\documentclass[lettersize,journal]{IEEEtran}

\usepackage{cite}
\usepackage{amsmath,amssymb,amsfonts}
\usepackage{algorithmic}
\usepackage{graphicx}
\usepackage{textcomp}
\usepackage{xcolor}

\usepackage{soul}
\usepackage{hyperref}
\usepackage{cleveref}
\crefname{section}{Section}{Sections}
\crefname{appendix}{Section}{Sections}
\crefname{figure}{Fig.}{Figs.}
\Crefname{figure}{Figure}{Figures}
\crefname{table}{Table}{Tables}
\crefname{equation}{Eq.}{Eqs.}
\Crefname{equation}{Equation}{Equations}
\crefname{algorithm}{Alg.}{Algs.}
\Crefname{algorithm}{Algorithm}{Algorithms}

\usepackage{graphicx}
\usepackage{textcomp}
\usepackage{multirow}
\usepackage{caption}
\usepackage{subcaption}
\usepackage{tikz}
\newcommand*\circled[1]{\tikz[baseline=(char.base)]{
            \node[shape=circle,draw,inner sep=0.5pt] (char) {#1};}}

\usepackage{pifont}

\usepackage{xspace}
\newcommand{\ourtech}{LightFAt\xspace} 

\usepackage{acro}
\DeclareAcronym{IPC}{short=IPC,long=instructions per cycle}
\DeclareAcronym{RA}{short=RA,long=Remote attestation}
\DeclareAcronym{ML}{short=ML,long=machine learning}
\DeclareAcronym{CFG}{short=CFG,long=Control-Flow Graph}
\DeclareAcronym{CFA}{short=CFA,long=Control-Flow Attestation}
\DeclareAcronym{RoA}{short=RoA,long=Region of Attestation}
\DeclareAcronym{OSVM}{short=OSVM,long=One-class Support Vector Machine}
\DeclareAcronym{LOF}{short=LOF,long=Local Outlier Factor}
\DeclareAcronym{IFO}{short=IFO,long=Isolation Forest}
\DeclareAcronym{PMU}{short=PMU,long=Performance Monitor Unit}

\def\BibTeX{{\rm B\kern-.05em{\sc i\kern-.025em b}\kern-.08em
    T\kern-.1667em\lower.7ex\hbox{E}\kern-.125emX}}

\usepackage[absolute]{textpos}
    
\begin{document}
\begin{textblock}{15}(0.5,0.3)
\noindent \large \centering
This is the author’s version of the work. The definitive  version will appear in the 2024 IEEE International
Symposium on Hardware Oriented Security and Trust (HOST)
\end{textblock}

\title{\ourtech: Mitigating Control-flow Explosion via Lightweight PMU-based Control-flow Attestation\\

\thanks{This work was partially funded by the Deutsche Akademische Austauschdienst (DAAD), the ``Helmholtz Pilot Program for Core Informatics (kikit)'' at Karlsruhe Institute of Technology, and the German Federal Ministry of Education and Research (BMBF) through the Software Campus Project.}
}

\author{\IEEEauthorblockN{Jeferson Gonzalez-Gomez\IEEEauthorrefmark{1}\IEEEauthorrefmark{2}, Hassan Nassar\IEEEauthorrefmark{2}, Lars Bauer, Jörg Henkel\IEEEauthorrefmark{2}}\\

\IEEEauthorblockA{\IEEEauthorrefmark{1} \textit{Instituto Tecnol\'ogico de Costa Rica (TEC)}}

\IEEEauthorblockA{\IEEEauthorrefmark{2}
\textit{Karlsruhe Institute of Technology (KIT), Chair for Embedded Systems (CES)} \\
\{jeferson.gonzalez, hassan.nassar, henkel\}@kit.edu}

}

\maketitle
\thispagestyle{empty} 

\begin{abstract} 
With the continuous evolution of computational devices, more and more applications are being executed remotely. 
The applications operate on a wide spectrum of devices, ranging from IoT nodes with low computational capabilities to large cloud providers with high capabilities. 
Remote execution often deals with sensitive data or executes proprietary software. 
Hence, the challenge of ensuring that the code execution will not be compromised rises. 
Remote Attestation deals with this challenge. 
It ensures the code is executed in a non-compromised environment by calculating a potentially large sequence of cryptographic hash values. 
Each hash calculation is computationally intensive and over a large sequence the overhead becomes extremely high.
In this work, we propose \ourtech: a \textbf{\underline{Light}}weight Control-\textbf{\underline{F}}low \textbf{\underline{At}}testation scheme. 
Instead of relying on the expensive cryptographic hash calculation, \ourtech leverages the readings from the processor's Performance Monitor Unit (PMU) in conjunction with a lightweight unsupervised \ac{ML} classifier to detect whether a target application's control flow is compromised, hence improving the system’s security. 
On the verifier's side, \ourtech reaches a detection accuracy of over 95\%, with low false-negative and false-positive rates.

\end{abstract}

\begin{IEEEkeywords}
security, attestation, machine learning, control flow
\end{IEEEkeywords}

\section{Introduction} \label{sec:intro}

Trust is an essential component in the current computing landscape.
Devices are becoming increasingly interconnected to communicate and cooperate on different computing applications. 
Among these applications are artificial intelligence applications, processing at the edge, and industrial automation~\cite{survey2022}.
All these applications would share sensitive data and the behavior of one device might be affected by the results of the computation on the other devices.
Thus, to ensure that no malicious activity would come from the remote computing parties, trust has to be established.

\ac{RA} enables the establishment of trust between two computing parties.
For example, users offload the computation and processing of their data to the cloud, and \ac{RA} allows them to make sure that the computation is not tampered with and is running as intended~\cite{survey2022}.
In this example, \ac{RA} works between a verifier (the user) and a prover (the remote computing device).
Usually, \ac{RA} occurs upon a request from the verifier to the prover in the form of a challenge-response protocol.
The verifier sends a challenge to the prover, which --based on the challenge-- calculates a cryptographic hash over a certain aspect of the system, to corroborate its normal functionality.

\ac{RA} can be either static or dynamic~\cite{survey2022}.
When static, the goal of the attestation is to ensure the integrity of the code memory and to ensure that the executed binaries are untampered.
In such a case, the state-of-the-art methods calculate a cryptographic hash over a certain range of code memory.
To avoid replay attacks, where the attacker records the value of a previous attestation and re-sends it, an initial value is sent to be concatenated with the memory range to ensure that two different attestation requests of the same memory range lead to two different responses.

Static attestation does not cover all the possible attacks.
Thus, dynamic attestation of the control flow, or \ac{CFA}, is performed~\cite{survey2022,cflat2016,ncallSlides}.
\ac{CFA} covers the different control-flow hijack attacks where an attacker tries to manipulate the control flow to reveal sensitive data or execute malicious code.
The typical way to perform \ac{CFA} is to calculate a sequence of cryptographic hash values over the sequence of executed \ac{CFG} nodes~\cite{cflat2016}.

\begin{figure} [t]
    \centering
    \includegraphics[width=0.6\linewidth]{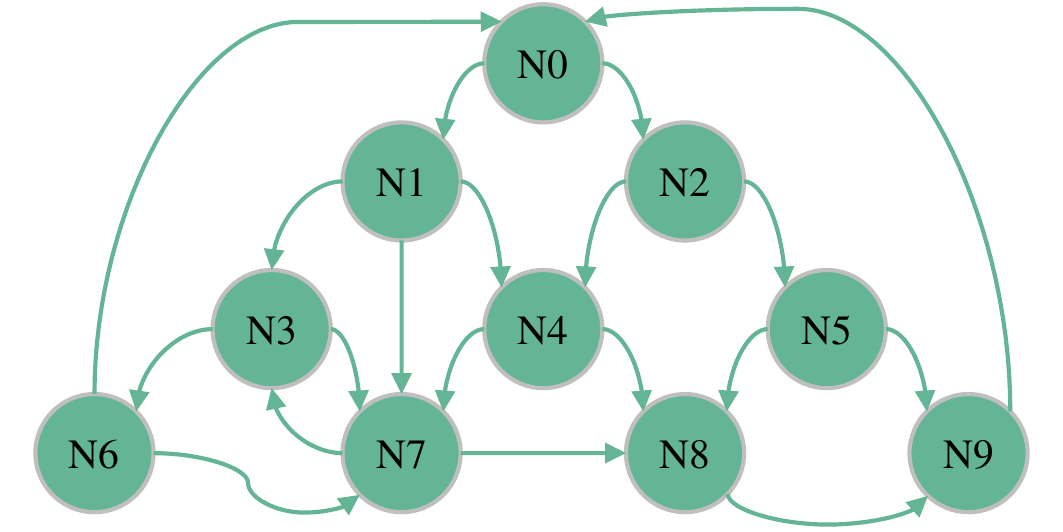}
    \caption{Example for a Control Flow Graph. With the several jumps and loops keeping track of each possible path becomes impractical.}
    \label{fig:cfg_expl}
\end{figure}

While this sequence of hashes would capture any deviation from the control flow, calculating a hash value with each executed node of the code leads to an overly expensive overhead that can range from hundreds to thousands of times increase in the execution time for applications with complex control flow~\cite{execTrace21}.
Thus, \textbf{the need for a lightweight \ac{CFA} solution is pressing}, for which we introduce our solution \ourtech: a novel unsupervised \ac{ML}-based attestation solution using the dynamic behavior of the application as an indicator of abnormal or malicious execution due to control-flow deviations. 
The \textbf{main contributions} of this paper are the following:
\begin{itemize}
    \item We are the first to propose employing execution readings from the CPU's \ac{PMU}, such as \ac{IPC} and L1 cache accesses, as indicators of normality in a control-flow attestation scheme.
    \item We employ a novel unsupervised \ac{ML}-based solution to control-flow attestation, which greatly reduces the overhead when compared to traditional solutions. 
\end{itemize}

\section{Background}
\label{sec:back}

\subsection{Remote Attestation}
\acf{RA} occurs between two types of parties: verifier and prover.
Using a challenge-response protocol, the verifier will be able to determine any tampering with the state of the prover. 
If the response from the prover does not match the expected value at the verifier, the verifier would label the prover as infected and stop communication with it~\cite{healed,shela,survey2022,wise}.
Addressing the infection, i.e., by stopping or fixing it, is out of the scope of \ac{RA}.
The verifier is usually assumed to have large computation capabilities and is trusted by the prover.

\subsection{Control Flow Explosion}
\label{sec:cfe}
\ac{CFA} suffers from control flow explosion~\cite{lape2020,lofat2017}.
Most of the \ac{CFA} solutions are only tested on small proof-of-concept applications.
However, real applications can have a variety of input dependencies and execution paths.
For example, \cref{fig:cfg_expl} shows a CFG that includes multiple loops and branch possibilities.
For the verifier, to keep track of all possible outcomes, it needs to calculate the cryptographic hash for each possible path, incl.\ each possible number of loop iterations.
Note that \cref{fig:cfg_expl} only shows a small snapshot.
For the whole program, the number of possible paths grows exponentially.

Some solutions try to deal with the control flow explosion~\cite{litehax,recfa,tinycfa}.
They mainly ignore loops and deal with them as one basic block.
However, attacks such as increasing or decreasing the number of loop executions or any attack within the loop will be undetected as the number of executions is not traced.
Another solution is to use the tracing capabilities of the prover, such as IntelPT for Intel-based provers, to evaluate whether the code is executed as intended or not.
However, such a solution on average incurs up to $36\times$ performance overhead and in the worst case might reach an overhead of $1000\times$ on the prover side, when performing attestation of an application with a complex control flow~\cite{execTrace21}. 

\subsection{Performance Monitor Unit (PMU)}
A \ac{PMU} is a hardware module that gathers various metrics or statistics from the dynamic operation of the processor and memory systems~\cite{arma53}.
A PMU is normally attached to the processor through an interface, records several events, and stores them in the Performance Counter registers.
These registers are available for readings through the operating system.
\acp{PMU} are commonly present in modern CPU architectures. 
In the security area, several works have leveraged the \ac{PMU} readings as attack exploits~\cite{spisak2016hardware, pmuleaker2023}, but also as assistance for detection and countermeasure mechanisms~\cite{timewarp2012,  gonzalezcache2023, dotecca,}.

\subsection{Machine Learning and Remote Attestation}

One relatively unexplored area is the feasibility of using Machine Learning (ML) for attestation~\cite{survey2022}.
Few examples exist in the literature, most of which deal with IoT scenarios~\cite{IoTrust,mlIoT}.
The idea is mainly that by comparing the behavior of the IoT nodes, the model will be able to detect if one acts oddly.

Another solution is proposed in~\cite{toqeerML} and deals with runtime attestation.
System calls of the prover are constantly monitored and fed to an ML classifier. 
If the prover is under attack, the system calls will be different and the ML classifier will be able to detect it.
To the best of our knowledge, there is no ML-based attestation solution that targets \ac{CFA}

\section{Our Lightweight Control-flow Attestation}
\label{sec:methodology}

\subsection{Target System, Threat Model, and Assumptions}

Current devices, even embedded ones, have become increasingly advanced and do not usually run bare metal applications anymore.
Modern systems such as Raspberry Pi and MPSoC FPGAs are capable of running fully functional operating systems.
This complicates the system and the computation running on it.
We target \ourtech for such systems with an operating system, not a bare metal one.
Our solution leverages the existing CPU's \ac{PMU}, commonly available in modern architectures for such systems \cite{spisak2016hardware}.

Our goal is to design an effective yet lightweight \ac{CFA} solution.
It should be lightweight enough to be used frequently to combat sneaky adversaries 
such as Time-of-Check to Time-of-Use (ToCToU) attacks~\cite{Kohnhauser2019}.
It should also be easy to use and train by application designers and achieve high accuracy with few to no false positives or false negatives.
Note that since we focus on the attestation scheme itself for a single device, IoT Swarm attestation~\cite{lisa}, physical attacks~\cite{darpa}, and restoring the state of the verifier~\cite{healed} are all out of scope for this work.

We have the following assumptions that are similar to those used by the state-of-the-art~\cite{cflat2016,lofat2017,litehax,recfa}.
We assume that the attacker wants to hijack the control flow through code injection or reuse.
Moreover, the attacker will not change the code memory content as we assume that \ourtech will run orthogonal to static attestation, which would catch any tampering with the memory code.
We also assume that the administrator of the system is trusted and consequently the code executing our attestation is trusted.

Unlike other hardware-assisted attestation schemes, \ourtech does not require any additional special hardware, i.e., it can be deployed on existing systems without further hardware modifications.

\subsection{\ourtech Attestation Flow} \label{sec:flow}

\begin{figure} [htp]
    \centering
    \includegraphics[width=\linewidth]{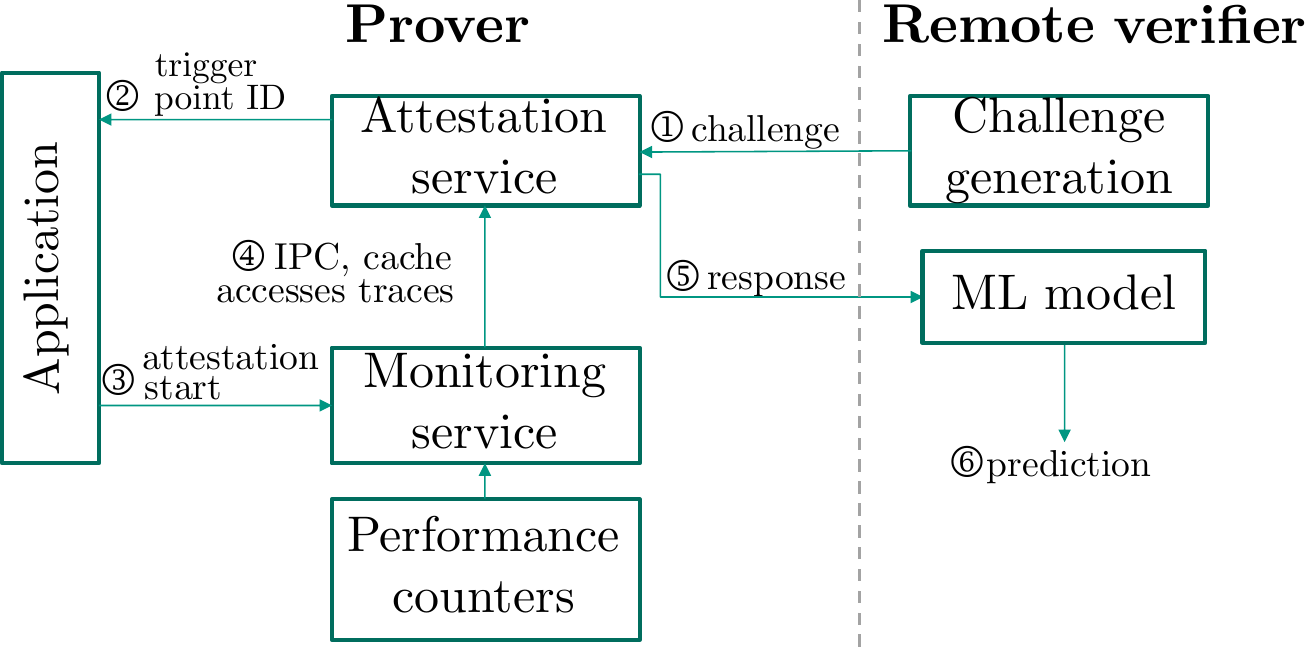}
    \caption{Overview of the \ourtech attestation flow including the prover and the remote verifier entities}
    \label{fig:high_level}
\end{figure}

\ourtech follows an attestation flow that is similar to a regular \acf{RA} scheme.
As represented in \cref{fig:high_level}, our attestation flow starts with the challenge~\circled{1} generation on the verifier side.
The challenge is received on the prover by a trusted attestation service, which translates the challenge into a trigger point.
The trigger point specifies the starting point of the monitoring for the performance (\ac{IPC}) and cache accesses metrics.
The attestation service specifies an trigger point identifier ID~\circled{2}, so the application knows where to start monitoring.
After receiving the trigger point ID, the application executes normally until it reaches the trigger point.
When the trigger point is reached, the application indicates the start of the attestation by generating a signal~\circled{3} to a trusted monitoring service.
This service reads the performance counter information from the application process and all its corresponding threads.
When the length of the monitoring window is reached, the monitoring service returns the \ac{IPC} and cache accesses traces~\circled{4} for the duration of the attestation to the attestation service.
Then, the attestation service signs the collected traces and sends them back to the verifier as the attestation response~\circled{5}.
The response signature is checked by the verifier to ensure it is untainted, and then the \ac{IPC} and cache access information is passed to an unsupervised \ac{ML}-based model, which emits the final verification as a prediction result~\circled{6}.
Note that \ourtech requires the code to be instrumented to add the trigger points.
This instrumentation can be done either by the application designer on the code level or the binary level by the verifier using a similar solution to~\cite{tinycfa,cflat2016}.
For the purposes of implementation and further evaluation, we build our solution to work on a Linux-based system.

\subsection{\ac{ML}-based Remote Attestation} \label{sec:mlverifier}

\begin{figure} [htp]
    \centering
    \includegraphics[width=\linewidth]{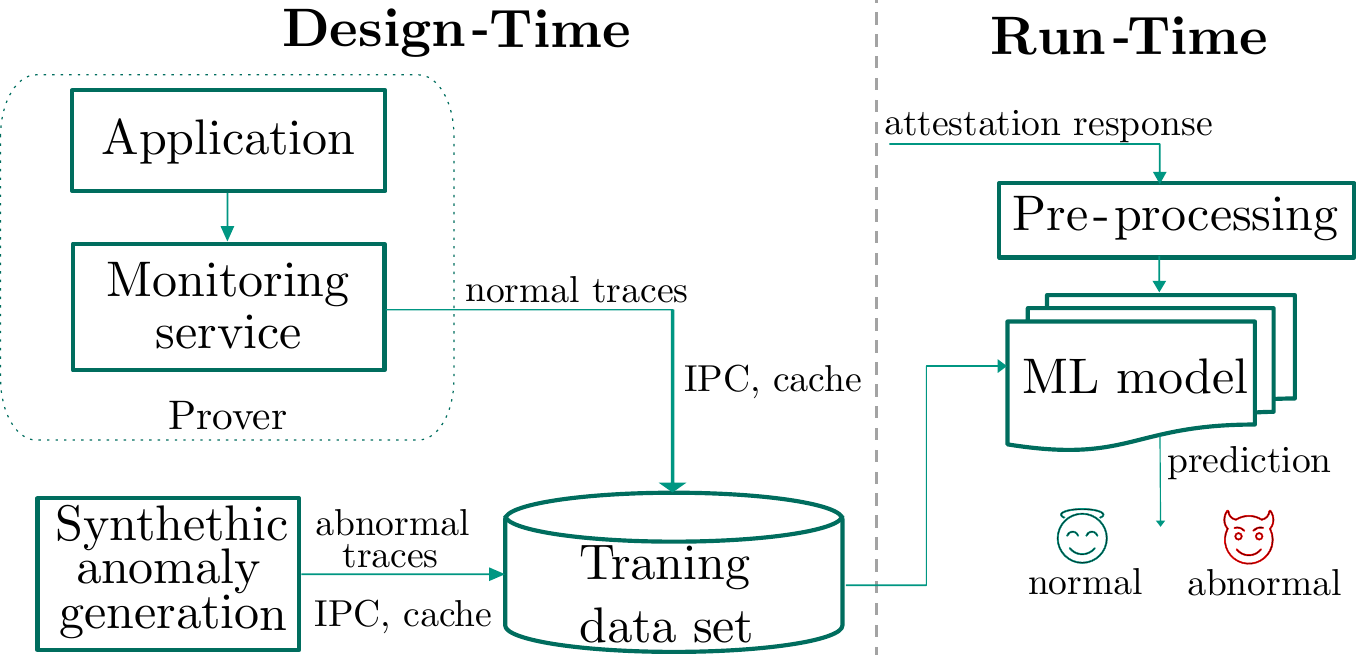}
    \caption{Overview of the ML-based verifier in \ourtech}
    \label{fig:ob_verifier}
\end{figure}

The main component of our solution is the \ac{ML}-based classifier that predicts whether the attestation response data (i.e., the \ac{IPC} and cache accesses information) corresponds to a normal or an abnormal execution.

One of the main challenges when dealing with an \ac{ML}-based solution to classify between two or more categories is the possible lack of training data.
In our case, collecting the normal data traces is rather straightforward, as it only requires measuring the different monitored executions.
However, abnormal traces are harder to get. 
First, since the application designers are not aware of vulnerabilities in their code (otherwise they would be fixed), modeling an attack to gather abnormal samples is not an easy task.
Second, even if an attack is modeled, collecting abnormal samples on a particular application will not reflect all possible abnormal executions, since the dynamic behavior of the attacker or abnormality is unpredictable and the possibilities of what it can do are vast.
To deal with this challenge, we chose to use the unsupervised learning technique, where models are fitted with mostly normal (unlabeled) execution traces, avoiding the need for an attack model or implementation.

The classification of the attestation response by the verifier's model can be seen as a two-part process, as depicted in~\cref{fig:ob_verifier}.
At design time, we compute the average of the \ac{IPC} and cache access window traces from runs of the application and create a training dataset of normal execution traces extracted from the monitoring service on the prover.
Each entry in the set contains the average of the IPC and the average of the cache accesses.
Note that for the design and training of the verifier, no real attack traces are needed.
However when evaluating our model in \cref{sec:model:eval}, we implement and use real attacks on a vulnerable application to validate the effectiveness of our solution.

At runtime, once the model has been trained, \ourtech gets the attestation response, computes the average for IPC and cache accesses, and feeds both features to the model to obtain the final prediction.

\section{Evaluation}
\label{sec:eval}

\subsection{Custom Vulnerable Application} \label{sec:motiv}

\begin{figure} [htbp]
    \centering
    \includegraphics[width=\linewidth]{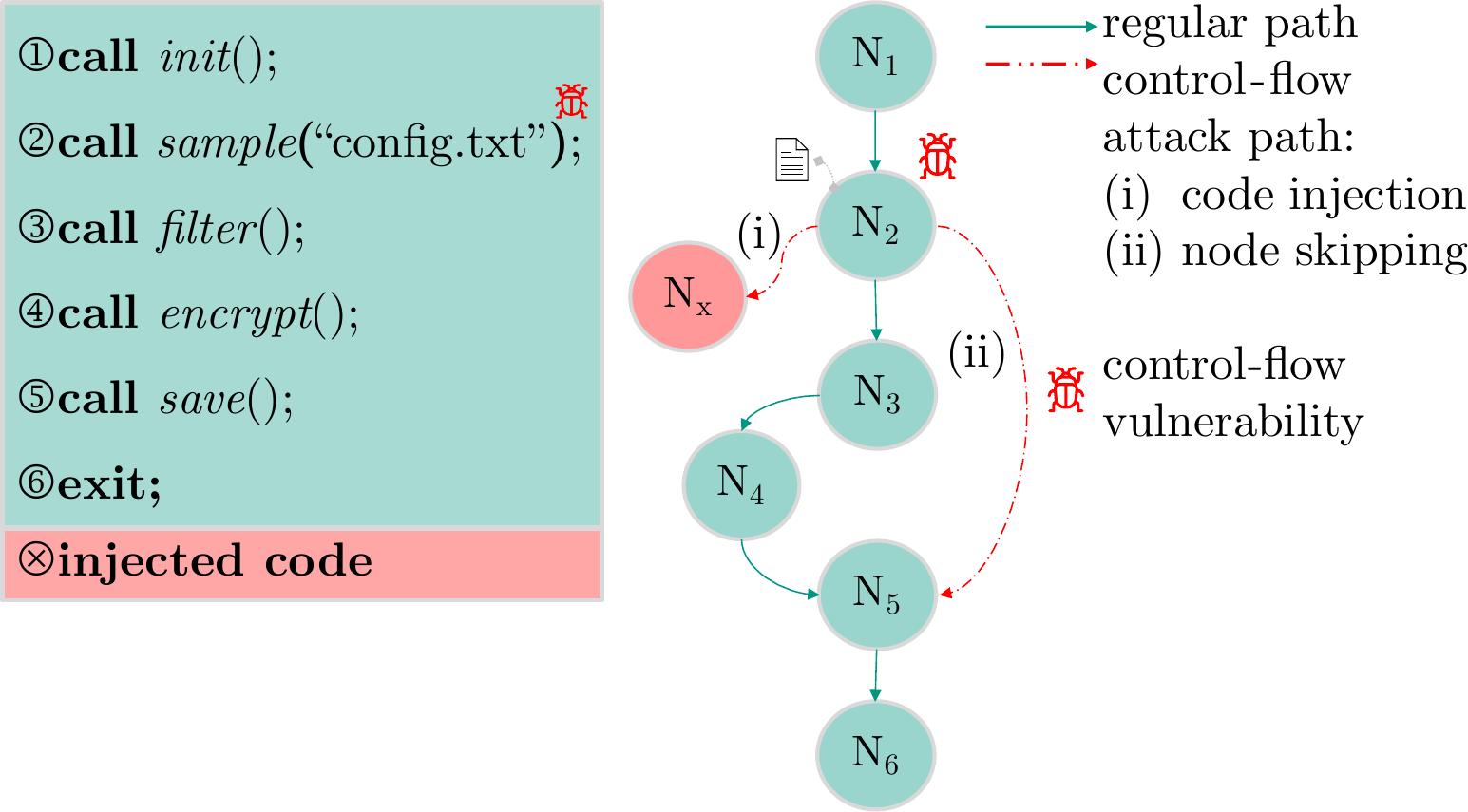}
    \caption{Overview of the control-flow attack paths on a vulnerable custom application}
    \label{fig:motiv_over}
\end{figure}

To evaluate \ourtech, we need a vulnerable application.
The attacker exploits this vulnerability to maliciously alter the control-flow of the application.
We created a custom application, depicted as an overview in \cref{fig:motiv_over}.
This application is meant to reflect an IoT-based monitoring application for user-sensitive data (e.g., blood pressure, heart beat, electrocardiograms, etc.).
After an initialization stage ($N_1$), the custom application enters a sampling stage ($N_2$), where it takes samples from user-sensitive sensors.
In this stage, the application first reads an unprotected configuration file (\textit{config.txt}) containing the user-defined monitoring parameters.
Then, it writes the content of the file to a buffer and applies the configuration before sampling the sensors.
After the sampling, the application enters a filtering stage ($N_3$).
When the filtering is done, the user's private data samples are encrypted ($N_4$) and then saved to a  file ($N_5$), before exiting ($N_6$).

However, this application exhibits a buffer overflow vulnerability in $N_2$, because the content of the input file is written directly into the buffer without any size checking.
This can be exploited by an attacker through a crafted payload written in the file to deviate the control flow of the application and execute arbitrary code ($i$) or perform node-skipping (e.g., to leak sensitive information) ($ii$).

Using the vulnerable application as described above, we create an attack where we exploit the vulnerability for both the code injection and data leakage scenarios.

\subsection{Experimental Setup and Data Collection} 

To evaluate \ourtech, we use monitor traces from running applications on a $3.7$\,GHz AMD Ryzen 7 2700X 64-bit processor, under CentOS7 as the operating system.
We instrument the custom vulnerable application on the code level with a trigger point for the attestation.
The monitoring for the trigger point runs for 300\,ms with a monitoring rate of 1\,ms. 
We run the application 1500 times with different inputs and parameters to get different possible behaviors.
We then run it another 1500 times but with attacks implemented.
We collect the IPC and cache access traces to feed them to the classifier.
For the classifier, We use the novelty detection unsupervised \ac{LOF}.
The model only learns the normal behavior and hence does not rely on the knowledge of any particular attack beforehand. 
For the training of the models, we use the scikit-learn Python library~\cite{scikit-learn}.

\subsection{\ac{ML} Model Evaluation} \label{sec:model:eval}

\begin{table}[htp]
\begin{center}
\caption{Performance of the classifier in differentiating the attack cases from the normal cases}
\label{tab:acc}
\begin{tabular}{cccccc}
\hline
Accuracy & FNR & FPR & Recall & Precision & F1 \\
\hline
95.62\% & 1.18\% & 5.61\% & 98.82\% & 96.40\% & 97.57\% \\
\hline
\end{tabular}
\end{center}
\end{table}

\Cref{tab:acc} shows the performance of our classifier, it reaches an accuracy higher than 95\%. 
Moreover, to better evaluate the effectiveness of the model, we measure the false positive rate (FPR), false negative rate (FNR), recall, precision, and F1 scores.
As can be seen from the table, the FNR (i.e., the ratio of attack traces being miss-classified as normal) is very low.
Although the FPR is higher than the FNR, the ratio is still less than 6\%
Moreover, as a security mechanism, \ourtech prioritizes minimizing the number of attacks that are miss-classified as normal, as this could have potentially grave consequences for the systems, which is not the case for false positives.

\subsection{Security and performance trade-off}
Our technique provides an interesting new trade-off.
Classical solutions can detect any deviations from the \ac{CFG}, as the sequence of hash computations must be unique for a specific execution path.
However, these solutions would incur in an exponentially increasing overhead by calculating a hash value with each node included in the path, which would be very large as well.
This means that the traditional attestation of applications with control-flow explosion might be unfeasible due to the effects that it would produce on the target system.
\ourtech, on the other hand, does not interrupt the code execution at all.
Rather, it only monitors the performance of the code.
Therefore, the overhead does not scale up as it does for other approaches.
In our experiments, the monitoring overhead for \ourtech was $1.26$\% for the evaluated application, which is insignificant, especially when compared similar approaches such as~\cite{execTrace21} with an overhead of $3.2\times$.
With this reduced overhead, our solution can be employed to increase the security of real-world complex systems by providing a lightweight attestation service that otherwise would be unfeasible due its complexity.
Although our method is not $100\%$ accurate, we are slightly compromising in terms of security with the benefit of massively reducing the overhead penalty thus allowing CFA even under control-flow explosion scenarios, which is not the case for most of the state-of-the-art solutions.

\begin{table}
\begin{center}
\caption{Execution time of the classifier on the verifier's side}
\label{tab:exec_mod}
\begin{tabular}{ccc}
\hline
  \begin{tabular}[c]{@{}c@{}}\textbf{Pre-processing}\\ \textbf{time} ($\mu$s)\end{tabular} &
  \begin{tabular}[c]{@{}c@{}}\textbf{Prediction}\\ \textbf{time} ($\mu$s)\end{tabular} &
  \multicolumn{1}{l}{\begin{tabular}[c]{@{}l@{}}\textbf{Total}\\ ($\mu$s)\end{tabular}} \\ \hline
   0.991 & 2.335 & 3.326 \\\hline
\end{tabular}
\end{center}
\end{table}

In addition to the low overhead incurred on the prover  side, \ourtech also reduces the complexity and computation cost on the verifier side.
As seen in \cref{tab:exec_mod}, the total execution time for our model is  very low. 
This has interesting implications for the implementation of real-time countermeasure to possibly compromised devices, which is not possible under other solutions, where the the verifier is expected to emulate all possible hash calculation sequences from all possible control-flow paths.
Instead, \ourtech is lightweight on the verifier side and it does not require any pre-computed CFGs.

\section{Conclusion}
\label{sec:conc}
In this paper, we presented \ourtech: an unsupervised \ac{ML}-based attestation scheme that leverages \ac{IPC} and cache access execution traces to predict whether the attestation response corresponds to normal or abnormal execution.
We have shown unsupervised \ac{ML} presents a high prediction accuracy of over $95$\%, with low false positive and false negative rates.
Our technique is lightweight and in contrast to state-of-the-art solutions it does not incur a linear overhead.
Therefore, it avoids the control-flow explosion problem as its monitoring behavior does not interrupt the execution of the code under attestation.

\bibliographystyle{ieeetr}
\bibliography{main}

\end{document}